\documentclass[a4paper,11pt]{article}
\usepackage{aaskaiid}
\newcommand{\apjl}{ApJL} 
\newcommand{\apjs}{ApJSS} 
\newcommand{\ao}{Appl. Opt.} 
\newcommand{\aap}{A\&A} 

\usepackage{aas_macros}%
\usepackage[utf8]{inputenc}
\hypersetup{
    colorlinks=true,
    linkcolor=red,
    filecolor=magenta,      
    urlcolor=magenta,
    citecolor=blue,
}
\usepackage{multirow}
\usepackage{orcidlink}
\usepackage{tabularx} 
\usepackage{adjustbox} 
\usepackage{pdflscape} 
\usepackage{graphicx} 
\usepackage{lscape} 
\usepackage{longtable} 
\usepackage{booktabs} 
\usepackage{caption} 
\DeclareUnicodeCharacter{202F}{\,} 

\title{Solar energetic particles and 
their association with radio emissions }
\ShortTitle{Solar energetic particles}

\author[1,2]{Diana E. Morosan\orcidlink{0000-0002-8416-1375}}
\ShortName{Morosan et al.} 
\author[3]{Anshu Kumari\orcidlink{0000-0001-5742-9033}}
\author[1]{Immanuel Jebaraj\orcidlink{0000-0002-1349-3663}}
\author[4]{Eduard P. Kontar\orcidlink{0000-0002-8078-0902}}
\author[5]{Mugundhan V.\orcidlink{0000-0002-8416-1375}}
\author[6,7]{Ketaki Deshpande\orcidlink{0000-0001-6861-6328}}
\author[1]{Nina Dresing\orcidlink{0000-0003-3903-4649}}
\author[8]{Puja Majee \orcidlink{0000-0002-2711-2366}}
\author[3,9]{Divya Paliwal\orcidlink{0009-0001-7689-0084}}

\affiliation[1]{Department of Physics and Astronomy, University of Turku, 20014, Turku, Finland}
\emailAdd{diana.morosan@utu.fi}
\affiliation[2]{Turku Collegium for Science, Medicine and Technology, University of Turku, 20014, Turku, Finland}
\affiliation[3]{Udaipur Solar Observatory, Physical Research Laboratory, Dewali, Badi Road, Udaipur - 313001, Rajasthan, India}
\emailAdd{anshu@prl.res.in}
\affiliation[4]{School of Physics \& Astronomy, University of Glasgow, Glasgow G12 8QQ, UK}
\affiliation[5]{Dept. of Space Planetary Astronomical Sciences and Engineering (SPASE), IITK, Kanpur 208016, India}
\affiliation[6]{Center for mathematical Plasma Astrophysics (CmPA), KU Leuven, Celestijnenlaan 200B, 3001 Leuven, Belgium}
\affiliation[7]{Solar-Terrestrial Centre of Excellence (STCE), Royal Observatory of Belgium, Avenue Circulaire 3, 1180 Uccle, Belgium}
\affiliation[8]{National Centre for Radio Astrophysics, Tata Institute of Fundamental Research, Pune, India}
\affiliation[9]{Indian Institute of Technology, Gandhinagar, Gujarat-382355, India}

\abstract{Energetic particle populations are ubiquitous throughout the Universe. In our solar system, the most prominent sources of energetic particles are solar flares or collisionless shocks often driven by huge eruptions of magnetised plasma called coronal mass ejections (CMEs). Remotely, low energy electrons from the Sun can be observed as solar radio bursts that are produced by accelerated electron beams undergoing beam-plasma interactions. There are still many open questions on the generation of solar energetic particles (SEP): how and where are SEPs accelerated during solar flares and CMEs and how they escape the solar atmosphere? Another important question is: what is the link between the solar radio bursts and the observed SEPs at spacecraft? SKA can provide high-resolution radio images combined with spectroscopic observations to determine the acceleration time, trajectory and escape of low energy electrons from the solar corona. The synergy between SKA and current space missions will help investigate solar activity and energetic particles across a wide range of wavelengths and particle energies. Particle data from spacecraft can be used to make a connection between radio bursts and SEPs by comparing SEP inferred injection times and energies to those of electrons generating radio bursts at the Sun. Radio observations in turn can be used to distinguish between flare and shock acceleration since different radio bursts pinpoint towards different energetic processes. Since the acceleration region and origin of SEPs of various properties is still largely debated, radio observations have the potential to be an invaluable tool in unraveling these processes.}

\begin{document}
\maketitle

\section{Introduction}

{Solar energetic particles (SEPs) are a common phenomenon in the heliosphere and can have energies ranging from a few keV to the GeV range. SEPs consist primarily of protons and electrons; however, heavier ions are also observed. SEPs in the heliosphere are produced by flaring processes on the Sun or shocks driven by coronal mass ejections (CMEs). Energetic electrons, in particular, are regularly detected remotely via the electromagnetic radiation in X-ray and radio wavelengths they generate through various processes, making them ideal to study in conjunction with remote observations. The relation between energetic electrons and other SEPs and the remotely observed X-rays and radio bursts is, however, still highly debated. This unclear relation makes it difficult to determine the origin of heliospheric SEPs at the Sun. The initial picture was that solar flares were considered to be the main mechanism producing solar energetic electrons, while shocks driven by CMEs were considered the main accelerators for protons \citep{lin1982, Reames1999}. However, there is increasing evidence indicating that the CME shock is also capable of accelerating energetic electrons \citep[e.g.][]{Dresing2022, Jebaraj2024,jebaraj24a, Jebaraj2023b,2024cosp...45..866K, morosan2025}.}

{Solar energetic electrons can be accelerated by flares during magnetic reconnection processes and they then travel both downward and upward from the acceleration region along magnetic field lines. Those electrons that propagate toward the dense chromosphere generate hard X-rays (HXR) through the bremsstrahlung mechanism \citep[e.g.,][]{krucker2011}, while the electrons that propagate outward frequently manifest as type III radio bursts. Type III bursts are generated via plasma instabilities following the propagation of an accelerated electron beam and they are co-temporal with the HXR emission, relating the upward and downward propagating electron populations \citep[e.g.,][]{vilmer2002, glesener2012, reid2014}. Another phenomenon that can accelerated solar energetic electrons are collisionless shocks driven by CMEs. In the low solar corona, electrons can be accelerated by near-perpendicular CME-driven shock waves to energies up to a few tens of keV \citep[][]{mann05,mann2022}. Following beam-plasma instabilities, the accelerated electrons can then generate electromagnetic emission at the plasma frequency and/or its harmonic \citep[e.g.,][]{wild1950, nelson1985, kl02, kumari2017a}. This emission can be detected remotely as type II radio bursts \citep[e.g.,][]{ne85, ma96, Kumari2023b} and is usually observed concurrently with type III bursts that are related to flare-accelerated electrons \citep[e.g.,][]{jebaraj2023,morosan2024, 2025arXiv251004586K}. Type II bursts are composed of numerous fine structures, the most well-known being herringbones, which are signatures of individual electron beams escaping the shock \citep[e.g.][]{holman1983,ca87,mann2018, mo19a, kumari2017b, magdalenic20}. \cite{2024A&A...683A.123Z} suggested that following advanced high spectral and temporal imaging spectroscopic observations of herringbones, they can originate from different processes and separate regions at the shock. Type II bursts and herringbones have since been found at multiple location around the shock and follow the expansion direction of CME shocks in the corona and interplanetary space \citep[e.g.,][]{zimovets2012, kumari2017c, zu18, mancuso19, mo19a, Morosan2020a, jebaraj2021, morosan2025b}. However, the full propagation of a type II radio bursts from the low to middle and higher corona is not well-known due to the limited frequencies that have been available for imaging with current instruments. Therefore, the possible escape path of type II generating electrons is also not very well-known and can vary from event to event. \citet{morosan2022} showed that herringbone bursts at frequencies above 150~MHz emanate from regions of exclusively closed field lines during the early stages of the CME eruption. However, a more recent study of a different event showed that, depending on the expansion direction of the CME, herringbones can also reach open field lines, further away from the eruption region \citep[][]{morosan2024}. The electrons generating herrignbones further away from the eruption site can potentially escape into the heliosphere. Energetic electrons observed by near-Sun spacecraft have not yet been studied in detail in conjunction with radio imaging. This is largely due to unfavorable spacecraft constellations relative to Earth, as well as the limited availability of radio imaging observations from the ground.   }

{In the case of energetic protons or ions, the detections are carried out in situ by spacecraft. A CME-driven shock is believed to be the main source of acceleration, especially in the case of relativistic SEPs \citep{2008AIPC.1039..190C,gopalswamy2014, 2015A&A...580A..80K, 2020arXiv200110506M, 2022Univ....8..275M,gopalswamy2024solarenergeticparticleevents}. High-energy protons correlate well with high-energy electrons \citep[e.g.][]{dresing22}, and as a result relativistic electrons observed at Earth are used as a method to provide lead times to proton radiation storms \citep{2007SpWea...5.5001P}. Type II radio bursts and solar energetic protons are also often related as they are common phenomena during fast and wide CME eruptions \citep{Kahler2019}. However, the origin at the Sun of high-energy SEPs is unknown and thus, it is also unknown if the lower-energy electrons (such as the ones producing type II bursts) and the energetic protons have a common origin.}

\section{Currently operational observatories monitoring solar radio activity and SEPs}

\subsection{Ground-based observatories}
\label{section2.1}

Ground-based observations of solar radio emissions are carried out using a range of spectrographs and radio interferometers that together provide broad frequency coverage and complementary diagnostic capabilities. Spectrographs primarily capture the spectral and temporal evolution of solar radio bursts, while interferometers deliver crucial spatial information, including source locations and morphologies — essential for understanding the relationship between the acceleration region of SEPs and their associated radio-emitting regions. A summary of the frequency coverage and operational period of the current ground-based instrumentation is available in Fig.~\ref{fig:fig2} and the daily observation windows of these instruments is shown in Fig.~\ref{fig:fig3}.

Prominent spectrographs include the Radio Solar Telescope Network (RSTN), the Observations Radiospectrographiques pour FEDOME et l’Etude des Eruptions Solaires \citep[ORFEES;][]{ORFEES2021}, the Yamagawa solar radio spectrograph \citep[YSRS;][]{Yamagawa2017}, the Gauribidanur Low-frequency Solar Spectrograph \citep[GLOSS;][]{gloss2014}, the Gauribidanur RAdio Spectro Polarimeter \citep[GRASP;][]{2023ApJ...958..181K}, the Nançay Decameter Array \citep[NDA;][]{ke97} and the global e-CALLISTO network \citep{2013EGUGA..15.2027M} operating in spectrographic/spectro-polarimetric mode. These instruments are vital for continuous monitoring and for conducting statistical analysis of solar radio emissions across a wide frequency range.

\begin{figure}[!ht]
    \centering
	\includegraphics[width=\columnwidth]{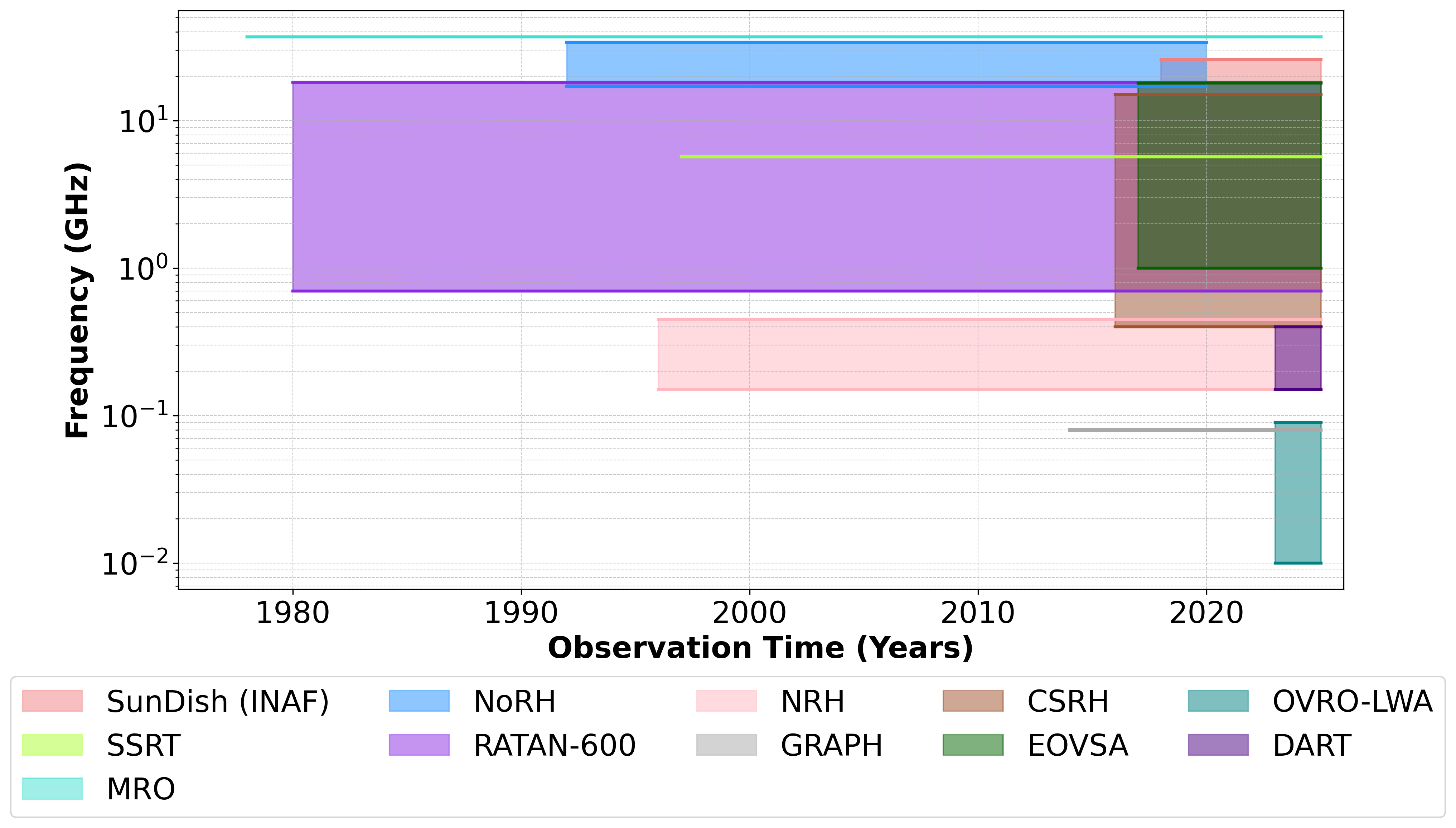}
    \caption{ Frequency coverage and operational periods of major solar radio imaging instruments worldwide.
    The plot illustrates the long-term continuity and complementary frequency ranges of facilities from metric to microwave wavelengths 
    Together, these observatories provide multi-frequency imaging coverage of the solar atmosphere over several decades, enabling studies of radio bursts, active regions, and coronal dynamics across multiple solar cycles.}
    \label{fig:fig2}
\end{figure}

At low frequencies (< 500~MHz), presently there are only a few solar dedicated interferometric imaging instruments such as the Gauribidanur Radioheliograph \citep[GRAPH;][]{1998SoPh..181..439R}, the Nançay Radioheliograph \citep[NRH;][]{ke97}, the Owens Valley Radio Observatory Long Wavelength Array \citep[OVRO-LWA;][]{chen2025}, the Daocheng Solar Radio Telescope (DART), and the Mingantu Spectral Radio Heliograph \citep[MUSER;][]{Yan2021}. Solar observations are also carried out by non-solar dedicated radio imaging instruments such as the LOw Frequency ARray \citep[LOFAR;][]{lofar2013}, the Murchison Widefield Array \citep[MWA;][]{Tingay2013}, and the Giant Metrewave Radio Telescope \citep[GMRT;][]{GMRT1991}. GRAPH and NRH have been operational for decades, producing solar images at frequencies between 50--450 MHz. New-generation arrays like LOFAR, MWA, and OVRO-LWA enable high temporal, spectral, and spatial resolution imaging of solar radio phenomena. LOFAR, a pathfinder for the Square Kilometre Array (SKA), and the MWA, a precursor to SKA-Low, have yielded key insights into the origin and mechanisms of various solar radio emissions \citep{reid17,mccauley2018,mo19a,2022SoPh..297..115L, 2025A&A...700A.274K, 2025arXiv250512932Z}.

{At higher frequencies (> 500~MHz), instruments such as the Expanded Owens Valley Solar Array \citep[EOVSA;][]{EOVSA2012}, the Siberian Solar Radio Telescope \citep[SSRT;][]{2003SoPh..216..239G}, and the upgraded Giant Metrewave Radio Telescope \citep[uGMRT;][]{uGMRT2017} have been used to study microwave and decimetric emissions in the few-gigahertz range. More recently, solar observations with MeerKAT \citep{jonas2016meerkat, Chen2021} have begun, marking a significant advancement in high-sensitivity, high-resolution studies of solar radio activity \citep{Kansabanik_2024, kansabanik_2025}
and solar flare electrons \citep{2026arXiv260205282L}. 
Importantly, it was demonstrated that inadequate spatial resolution in solar flare radio observations could lead to over-estimation of the number of accelerated electrons \citep{2023ApJ...947L..13K,2025ApJ...990..101V}, and making it hard to obtain a consistent picture of flare accelerated electrons required for X-ray and radio emissions.}

\begin{figure}[!ht]
    \centering
	\includegraphics[width=\columnwidth]{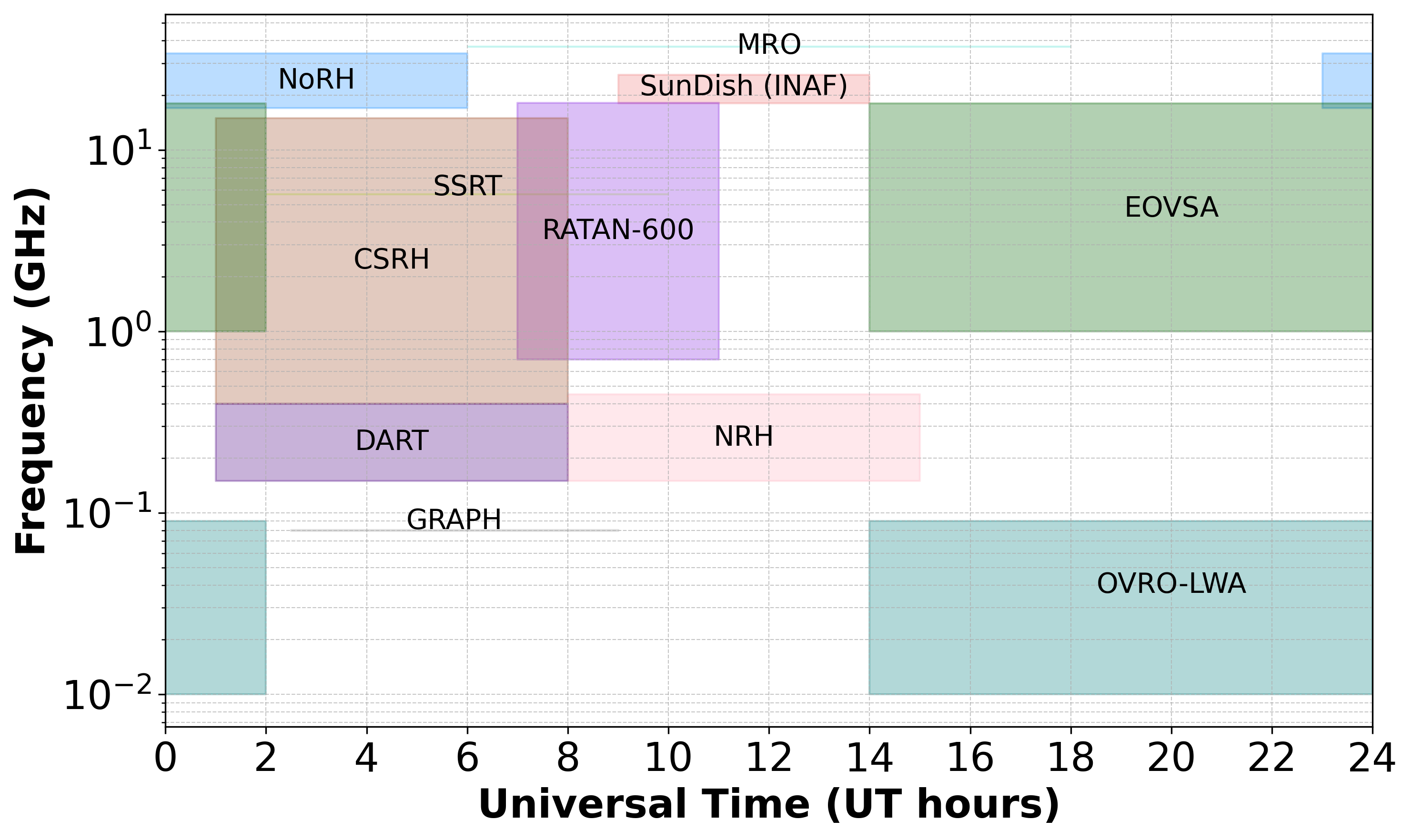}
    \caption{Typical daily observation windows and frequency coverage of solar radio imaging instruments.
    Each shaded region represents the frequency band and local observing time (in Universal Time) for facilities across different longitudes. The plot highlights the near-continuous global coverage achieved through coordinated observations from Asia (GRAPH, DART, CSRH), Europe (NRH, RATAN-600, SunDish, MRO, SRH), and the Americas (EOVSA), facilitating 24-hour monitoring of solar radio emissions from 0.05~GHz to 37~GHz. The figure shows the daily maximum observation window for each instrument which would correspond to summer time observations, however, these observation windows can vary. }
    \label{fig:fig3}
\end{figure}

\subsection{Space-based observatories}
\label{section2.2}

\begin{figure}[!ht]
    \centering
	\includegraphics[width=0.65\columnwidth]{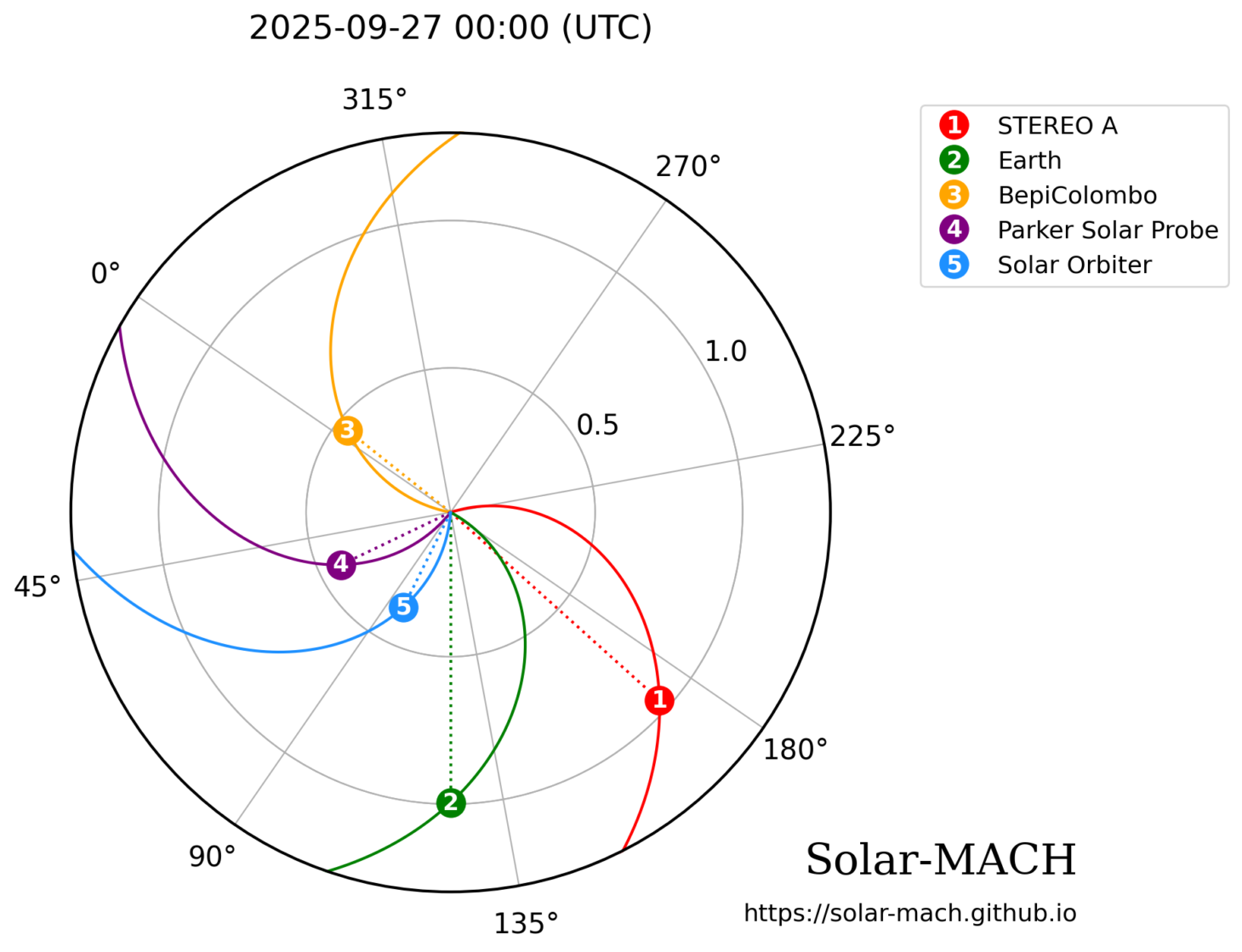}
    \caption{SolarMACH plot showing the current constellation of spacecraft monitoring the Sun \citep{gieseler2023}. Multiple spacecraft are also located close to Earth. The solid spirals show the magnetic connections of each spacecraft and the Earth to the Sun along the interplanetary magnetic field.  }
    \label{fig:fig1}
\end{figure}

{Solar activity and SEPs can now be monitored by an unprecedented fleet of spacecraft that contain multiple instruments for both remote and in situ observations. Of interest in solar activity studies, are remote observations by spacecraft primarily in white-light, extreme-ultraviolet (EUV) and X-ray wavelengths. Observations at these wavelengths provide information on the accompanying solar activity and they can be used to help distinguish the sources of SEPs, for example flares or shocks driven by coronal mass ejections that are described in more detail in Section 3. Most observing spacecraft are located close to Earth (see Figure~\ref{fig:fig1}), while STEREO-A, Solar Orbiter and Parker Solar Probe can be located at different positions in orbit around the Sun. Solar Orbiter and PSP are in orbits such that they can reach even closer distances to the Sun than the Sun--Earth distance. }

Monitoring solar activity from the Earth's view point is crucial for space weather forecasting. Solar transients such as CMEs, flares, and solar wind can be tracked both remote sensing and in-situ observations. Since the launch of Solar and Heliospheric Observatory \citep[SOHO;][]{do95}, space-based solar observations have improved drastically. SOHO provides coronagraphic observations from Earth’s perspective, which are highly valuable for tracking CME propagation. The Solar Dynamics Observatory \citep[SDO;][]{pe12} offers complementary EUV observations at different wavelengths, providing insights into flare locations and CME initiation sites. It also provides synoptic magnetograms that give information on the complexity of the active region and are used in different space weather models as an input parameter. The Geostationary Operational Environmental Satellite \citep[GOES;][]{Garcia94} provides continuous X-ray observations of the Sun offering information on solar flare intensity and their classification and also more recently EUV observations \citep{SUVI2022}. 

Along with these remote sensing observations, different spacecraft at the Langrangian point L1, such as Wind \citep{harten1995},  Deep space climate observatory \citep[DSCOVR;][]{Burt12}, and  Advanced Composition Explorer \citep[ACE;][]{stone1998}, provide in situ measurements of the solar wind plasma including magnetic field, velocity, density, temperature etc., that are used to track solar transients and solar wind structures. In situ measurements of energetic particles such as protons, electrons, alpha particles and heavy ions can be obtained by GOES, Wind, DSCOVR, ACE and SOHO.  

Observations of the Sun from different perspectives are also crucial for obtaining a complete picture of solar activity. The Solar TErrestrial RElations Observatory \citep[STEREO;][]{Kaiser2008} consists of two identical spacecraft in heliocentric orbits, namely STEREO-A and STEREO-B, with STEREO-B observations available until October 2014. It observes the Sun from various vantage points at a distance of ~1~AU. In addition to white-light and EUV observations, it also provides in situ measurements. However, these observations were limited to ~1~AU until the arrival of the Parker Solar Probe \citep[PSP;][]{psp2016} and Solar Orbiter \citep[SolO;][]{muller2020}. PSP and SolO offered new set of observations revealing  different aspects of solar activity. PSP traveled closer to the Sun than any other spacecraft and reached a distance as close as 6.1 million km ($\sim8.8~R_\odot$). It allowed mapping of the plasma parameters at various distances in the solar corona using onboard instruments such as FIELDS \citep{Bale16}, Solar Wind Electrons, Alphas, and Protons \citep[SWEAP;][]{Kasper2016} and Integrated Science Investigation of the Sun \citep[IS$\odot$IS;][]{mccomas2016}. SolO provided high resolution imaging observations with instruments like the Extreme Ultraviolet Imager \citep[EUI;][]{eui2020} and the Spectrometer/Telescope for Imaging X-rays \citep[STIX;][]{stix2020}, as well as in situ measurements offering valuable complementary information on solar activity from different distances and vantage points from Earth to the Sun. SolO also has an Energetic Particle Detector \citep[EPD;][]{Rodriguez-Pacheco2020} onboard, specifically dedicated to the study of SEPs, measuring protons, electrons, and heavy ions over a wide energy range (up to several hundred MeV/nucleon) and with high temporal and unprecedented energy resolution. Its different vantage point measurements provide insights on the SEP injection, acceleration and propagation through the heliosphere.

Tables~\ref{tab:table1} summarises the currently operational spacecraft that can be used in conjunction with SKA observations to investigate SEP events and their radio signatures. These missions collectively provide complementary measurements of CMEs, flares, X-rays, in situ electrons and ions across a wide energy range, and low-frequency radio bursts from kilometric to metric wavelengths. Table \ref{tab:table2} summarises the principal space-based radio instruments that provide dynamic spectroscopic observations of solar and interplanetary radio emissions. These instruments operate primarily at kilometric to decametric wavelengths, covering frequencies from a few kHz to several tens of MHz, and are essential for tracking Type II and Type III radio bursts associated with CMEs and solar energetic particle (SEP) events as they propagate through the heliosphere.

\begin{table}[!ht]
\centering
\caption{Summary of currently available spacecraft that can be used in conjunction with SKA to study SEPs. }
\begin{tabular}{l|l|ll}
\hline
\multicolumn{1}{c}{\textbf{Spacecraft}}                            & \multicolumn{1}{c}{\textbf{Observables}}                                                                                                                         & \multicolumn{1}{c}{\textbf{Operational Years}}       &  \\ \hline
\begin{tabular}[c]{@{}l@{}}Solar Heliospheric Observatory \\ (SOHO)\end{tabular}            & \begin{tabular}[c]{@{}l@{}}CMEs (white-light)\\ Protons (1.3-130 MeV)  at 1 AU\end{tabular}                                                                                                 & 1996 – present                                                        &  \\ \hline  
\begin{tabular}[c]{@{}l@{}}Solar Dynamics Observatory \\ (SDO)\end{tabular}                 & CMEs and flares (EUV) at 1 AU                                                                                                                                                               & 2010 – present                                                          &  \\ \hline  
\begin{tabular}[c]{@{}l@{}}Solar Terrestrial Relations \\ Observatory (STEREO)\end{tabular} & \begin{tabular}[c]{@{}l@{}}CMEs (EUV and white-light)\\ Radio bursts (10 kHz – 16 MHz)\\ Electrons (45 – 425 keV) \\ and other SEPs  at 1 AU\end{tabular}                                   & \begin{tabular}[c]{@{}l@{}}STEREO-A:\\ 2007 – present\end{tabular}      &  \\ \hline

Hinode  & Flares (X-rays) &  2006 – present     &  \\ \hline  

Solar Orbiter (SolO)                                                                        & \begin{tabular}[c]{@{}l@{}}Electrons (2 keV – 18 MeV) and \\ other SEPs at 0.3 – 1 AU\\ CMEs, flares (EUV, white-ligh)\\ X-rays (4 - 150 keV)\\ Radio bursts (10 kHz – 16 MHz)\end{tabular} & \begin{tabular}[c]{@{}l@{}}2021 – 2026\\ (nominal mission)\end{tabular} &  \\ \hline  
Parker Solar Probe (PSP)                                                                    & \begin{tabular}[c]{@{}l@{}}Electrons (25 – 1000 keV) \\ and other SEPs at 0.05 – 1 AU\\ Radio bursts (10 kHz – 19 MHz)\end{tabular}                                                         & \begin{tabular}[c]{@{}l@{}}2018 – present\\ (nominal mission)\end{tabular} &  \\ \hline  
BepiColombo                                                                                 & \begin{tabular}[c]{@{}l@{}}Electrons (60-3000 keV) \\ and other SEPs from 0.3 – 1 AU\\ X-rays (1 -- 20 keV)\end{tabular}                                                                       & \begin{tabular}[c]{@{}l@{}}2018 – 2026 \\ (cruise phase)\end{tabular}   &  \\ \hline  
Wind                                                                                      & \begin{tabular}[c]{@{}l@{}}Electrons (20 – 650 keV) \\ and other SEPs at 1 AU\\ Radio bursts (20 kHz – 13 MHz)\end{tabular}                                                                 & 1995 – present             &  \\ \hline
\begin{tabular}[c]{@{}l@{}}Advanced Composition Explorer \\ (ACE)  \end{tabular}                                                                                     & \begin{tabular}[c]{@{}l@{}}Electrons (45 – 315 keV) \\ and other SEPs at 1 AU\end{tabular}                                                                 & 1997 – present             &  \\ \hline
\begin{tabular}[c]{@{}l@{}}Geostationary Operational \\ Environmental Satellite \\ (GOES)     \end{tabular}                                                                                  & \begin{tabular}[c]{@{}l@{}}CMEs and flares (EUV) \\ X-rays (0.1 -- 0.8 nm) \\ Electrons (50 keV -- 4 MeV) \\ and other SEPs at 1 AU\end{tabular}                                                                 & 2016 – present (EUV)            &  \\ \hline
\begin{tabular}[c]{@{}l@{}}Aditya-L1\end{tabular} & 
\begin{tabular}[c]{@{}l@{}}CMEs and flares (EUV, three \\ emission lines of the corona)\\
X-rays (1 – 150 keV)\\
Protons and electrons (20 keV – \\ 5 MeV)
at L1 (1 AU)\end{tabular} &
2023 – present & \\ \hline

MAVEN  &  \begin{tabular}[c]{@{}l@{}}SEPs (20 keV – 6 MeV) \\ at 1.38 – 1.67~AU\end{tabular} &   2014 – present     &  \\ \hline  
\end{tabular}
\label{tab:table1}
\end{table}

\begin{table}[h!]
\centering
\caption{Summary of space-based radio instruments.}
\resizebox{\linewidth}{!}{%
\begin{tabular}{l|l|c|l|c}
\hline
\textbf{Spacecraft} & \textbf{Instrument} & \textbf{Freq. (MHz)} & \textbf{Type} & \textbf{Operational Years} \\ \hline
Voyager 1/2 & PWS & 0.00001--0.056 & Radio & 1977--present \\ \hline
Ulysses & URAP & 0.001--1 & Radio & 1990--2009 \\ \hline
Wind & WAVES & 0.02--13.825 & Radio Spectrometer + DF & 1994--present \\ \hline
STEREO A/B & SWAVES & 0.0025--16 & Radio Spectrometer + DF & 2006--present \\ \hline
PSP & FIELDS/RFS & 0.01--19.2 & Radio + Electric Field + DF &  2018--present \\ \hline
Solar Orbiter & RPW & 0.0005--16 & Radio + Electric Field + DF & 2020--present  \\ \hline
\end{tabular}%
}
\label{tab:table2}
\end{table}

While all these remote sensing observations and in situ measurements provide insights on the solar activity, radio observations offer a unique window into these processes. The ground-based radio observations cannot be conducted below approximately $\mbox{10\,--\,30\,MHz}$ due to the ionospheric cut-off. Low frequency radio signals below this cut-off are reflected back into space by the Earth's ionosphere, which makes space-based observations necessary. Therefore, space-based solar radio observations cover the lower frequency range of $\sim\mbox{0.01\,--\,20\,MHz}$. The instruments Wind/WAVES \citep{bougeret1995}, STEREO/WAVES \citep{bougeret200B}, PSP/FIELDS/Radio Frequency Spectrometer \citep[RFS;][]{rfs2017} and SolO/Radio and Plasma Waves \citep[RPW;][]{Maksimovic2020} provide spectroscopic observations in this frequency range. Different positions of these instruments are useful to obtain a general view on the propagation of solar energetic particles and the location of the solar radio bursts \citep{Musset21, Badman22}. Wind/WAVES and STEREO/WAVES also provide direction-finding (DF) observations at discrete frequencies in the range of about $\sim\mbox{0.1\,--\,2\,MHz}$, which can be used to obtain 3D radio source positions \citep{Reiner98, Krupar14b, Krupar14a, Magdalenic14, jebaraj2020, Deshpande2025}. RPW and RFS are also equipped with DF capabilities to provide similar observations.  

In addition to all existing space-based observatories, newly launched missions like Aditya-L1 \citep{Tripathi2023AdityaL1}, Polarimeter to Unify the Corona and Heliosphere \citep[PUNCH;][]{DeForest2025} and PROBA-3 \citep{Zhukov2025ASPIICS} provide new sets of observations. The instruments onboard Aditya-L1 such as Aditya Solar wind Particle EXperiment \citep[ASPEX;][]{Goyal2018ASPEX} provide measurements of a wide energy range of energetic particles (\mbox{100\,eV\,--\,20\,MeV}). In addition, it also provides both remote sensing observations and in situ measurements of the solar wind \citep[]{Tripathi2023AdityaL1}. The PUNCH mission consists four satellites which together provide large field of view imaging observations of the solar corona and heliosphere. These observations can be used to track solar transients such as CMEs over long distances and they also help to reconstruct the CME structure in 3D. PROBA-3 mission comprises of two spacecraft (a coronagraph and an occulter) which employs a precise formation flying technique to work as one spacecraft. The onboard payload ASPIICS \citep[Association of Spacecraft for Polarimetric and Imaging Investigation of the Corona of the Sun;][]{Zhukov2025ASPIICS} uses this technique to form a giant coronagraph which can observe the solar corona between \mbox{1.097\,--\,3\,$R_\odot$}. Combining these observations with already existing coronagraph observations give a complete view of the solar corona which is useful to understand the evolution of solar transients. All these new observations fill existing gaps in the solar and heliospheric studies and provide new insights on the underlying physical processes of solar activity.

\section{From remote sensing to in situ observations of solar energetic particles}
\label{section3}

{Remotely, the generation and motion of non-thermal electrons can be tracked in dynamic radio spectra in the microwave and metric-to-kilometric wavelength that cover processes starting in the corona and propagating into interplanetary medium. For example, Type~III bursts, produced by electron beams escaping along open magnetic field lines, provide a temporal signature of the initial particle release and its connectivity to the heliosphere \citep{reid2014}. Type~II bursts, generated by shock-accelerated electrons at CME-driven fronts, mark the presence and speed of propagating shock waves that are capable of accelerating ions observed later as SEPs \citep{CaneErickson2005}. A significant addition to dynamic spectra is interferometric radio imaging, which can pinpoint the exact location and propagation of electron beams following acceleration at the Sun \citep[for a comprehensive review of low frequency solar radio bursts see, for example,][]{Kumari01.2026.SKA}.}

{In the heliosphere, in situ observations of SEPs now span more than five decades, since the first detection of impulsive electron events originating from the Sun with energies $>40$~keV \citep{1965JGR....70.5737V}. In situ instruments onboard spacecraft can now provide particle fluxes, energy spectra, pitch angle distributions, and composition at fixed points in the heliosphere from a few tens of keV to GeV. Radio measurements add timing and geometric context by tracing where non-thermal electrons first escape and where shocks form and evolve in the corona and interplanetary medium \citep{cairns2003,Klein_Dalla2017}. By combining remote multi-wavelength and in situ observations, a complete picture can be constructed from the onset of particle acceleration all the way to their arrival at spacecraft.}

{The following subsections outline the importance of combined remote and in situ observations in determining the possible origins of SEPs:}

\subsection{Type II bursts and Solar Energetic Particles}
\label{section3.1}

{Solar energetic electrons are of particular interest because they emit electromagnetic radiation, enabling remote observations and allowing us to infer their acceleration mechanisms and acceleration sites. The CME shock is a great accelerator of SEPs \citep[e.g.][]{gopalswamy2014}, however, it is hard to pinpoint the contribution of the shock in accelerating low energy electrons, especially when a flare is almost always present during CME eruptions. Remote radio imaging of energetic electron beams can be further enhanced when these electrons are also measured in situ \citep[e.g.,][]{jebaraj2023, morosan2024}, as it provides insights into the origin and acceleration mechanism of the in situ electrons as well. This has only been possible recently, with the unprecedented fleet of spacecraft in the heliosphere monitoring SEPs. 
A positive correlations was also found between the bandwidth of the hectometric type II bursts and the peak flux of the SEPs, hinting that they may be accelerated by the same shock region \citep{iwai2020}.} 

{Recent studies demonstrating spatial connections between remote type II and in situ observations of electrons showed that escaping radio emitting electrons intersected magnetic field lines that were well-connected to spacecraft observing co-temporal electron events \citep[][]{morosan2024,morosan2025}. In the study of \citet{morosan2025}, low-energy electrons accelerated by the shock intersected field lines connecting to Solar Orbiter at 0.3~AU, which detected in situ electrons well correlated in time with the radio emission. Radio imaging can effectively demonstrate the role of shock acceleration in the electrons observed in situ by connecting the trajectories of the imaged type II radio sources with interplanetary magnetic field lines that connect to the in situ observers (see Figure~\ref{fig:fig2} adapted from \citealt{morosan2025}). However, the energy spectrum of the in situ electrons from the study of \citet{morosan2025} showed a spectral index that lies in between the indices expected from either flare and shock acceleration, even though the flare connection to open field lines was not as clear as that of the shock. Future observations of multiple energetic events are necessary to link the possible spectral shapes to the different types of radio bursts. Linking the occurrence of type II and type III radio bursts with magnetic connections to spacecraft can be used to better characterize the origin of in situ electrons and the flare and shock contributions to energization. Timing and bandwidth analyses can then also be carried out in relation to other SEPs (ions) to determine if they have a common origin \citep[e.g.][]{iwai2020}. }

\begin{figure}[!ht]
    \centering
	\includegraphics[width=1.09\columnwidth]{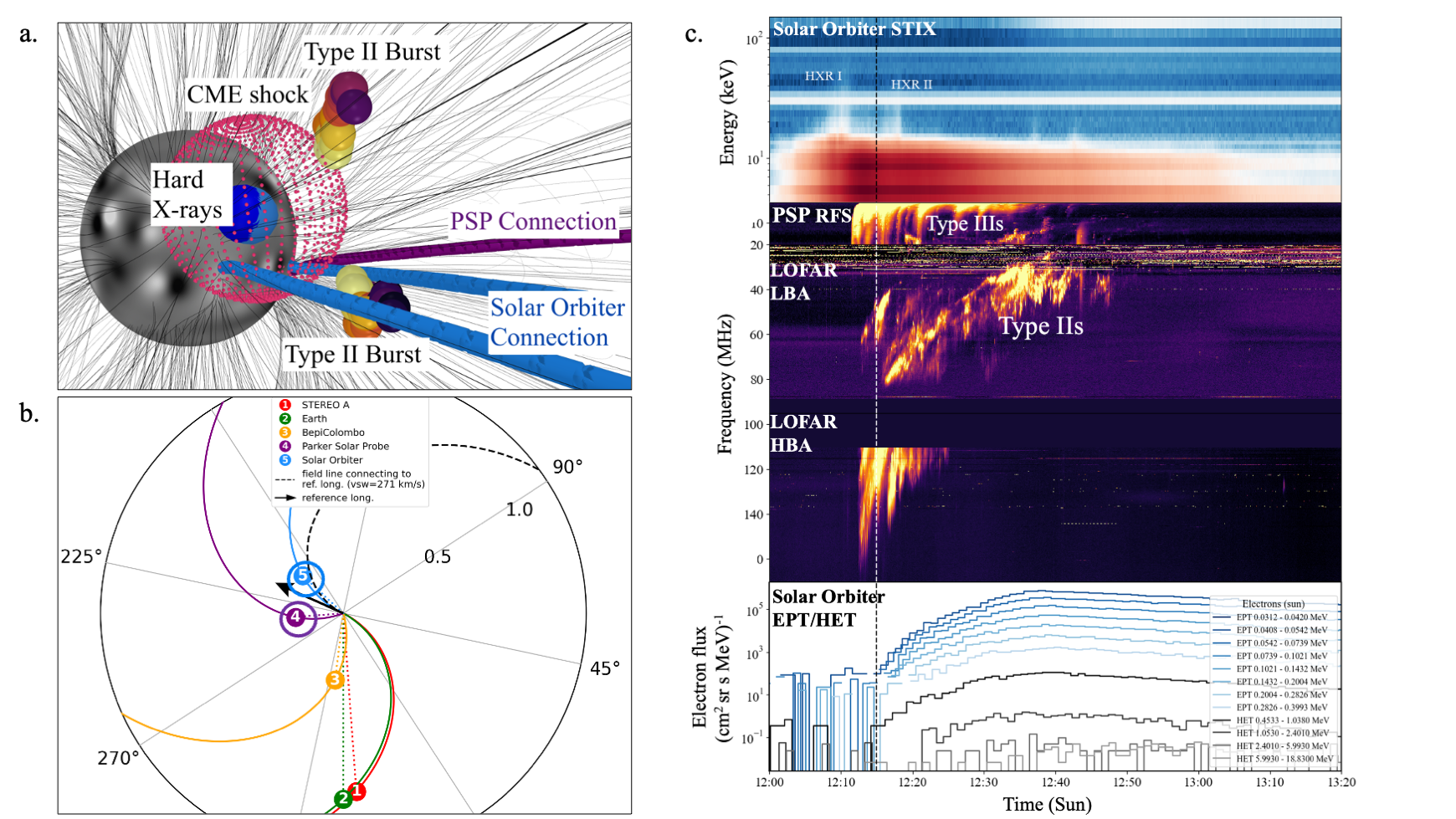}
    \caption{Connecting remote radio and X-ray observations to solar energetic electrons. a. Three-dimensional representation of the electron acceleration signatures (hard X-rays and Type II radio bursts) that occurred during a CME eruption of 3 October 2023. The purple mesh represents the CME shock outline and the purple and blue magnetic field lines show the magnetic connections to two spacecraft. b. Solar-MACH plot of the spacecraft configuration on 3 October 2023 where the black arrow represents the coordinates of the flare. c. X-rays, radio bursts and energetic electrons observed by Solar Orbiter/STIX, PSP/RFS and LOFAR, and Solar Orbiter/EPT/HET, respectively. The type II bursts shows a good temporal association with the in situ energetic electrons and it is also the remote feature that is best magnetically connected to Solar Orbiter. Figure adapted from \citet{morosan2025}.}
    \label{fig:fig2}
\end{figure}

\begin{figure}[!ht]
    \centering
    \includegraphics[width=0.95\columnwidth]{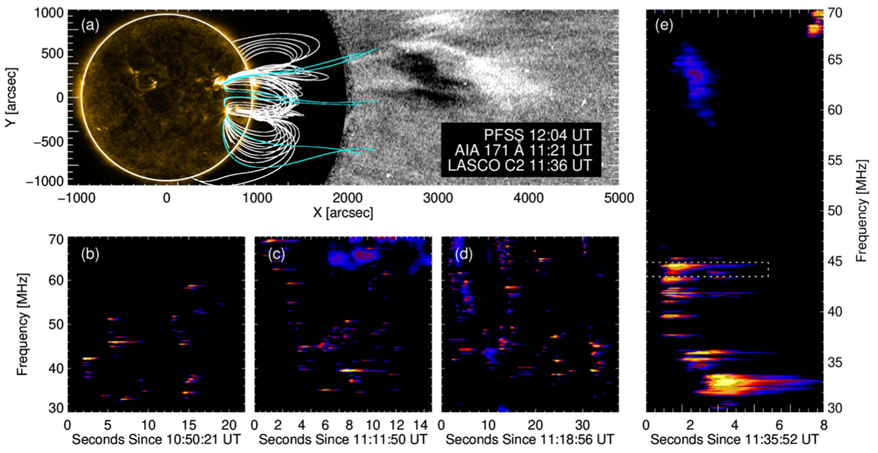}
    \caption{Coronal Mass Ejections (CME) and signatures of  accelerated electrons. \textbf{(a)} SDO/AIA $171$ \r{A} image at 11:20:57 UT, superimposed with a potential field source surface (PFSS) extrapolation at noon showing both open (blue) and closed (white) field lines within region surrounding the active region and northern sunspot, and a LASCO C2 image showing the streamer-puff (above) and narrow (below) CME fronts at 11:36:05 UT, as shown in \citep{2020ApJ...893..115C}. \textbf{(b-d)} Dynamic spectra showing samples of the spike emission. \textbf{(e)} Dynamic spectra of a Type IIIb J-burst. All dynamic spectra are background subtracted where the background is defined using a region at the start of each dynamic spectra containing no bursts at all frequencies. Image taken from \citep{2023ApJ...946...33C}.}
    \label{fig:cme_spikes} 
\end{figure}

\subsection{CMEs and non-shock electron acceleration}
\label{section3.2}
CME-driven shocks can accelerate electrons to tens of keV energies, which can generate type II radio bursts. In addition, CME flanks can also produce spike bursts \citep{2023ApJ...946...33C}, which are signatures of small-scale acceleration processes likely due to magnetic reconnection. These spike bursts occur in close temporal association with type II bursts \citep{2020ApJ...893..115C}, but the spatial location is different suggesting that the CME perturbs the magnetic field lines causing small-scale reconnection and electron acceleration. Figure~\ref{fig:cme_spikes} shows an example of spike bursts occurring during a CME eruption. The spike bursts occur during the expansion phase of the CME and are located at regions different to those of type IIs. The details of these small-scale acceleration processes that occur during CME eruptions are still not well understood. 

\subsection{Hard X-rays, Type III bursts and Solar Energetic Electrons}    
\label{section3.3}

\begin{figure}[!ht]
    \centering
	\includegraphics[width=0.8\columnwidth]{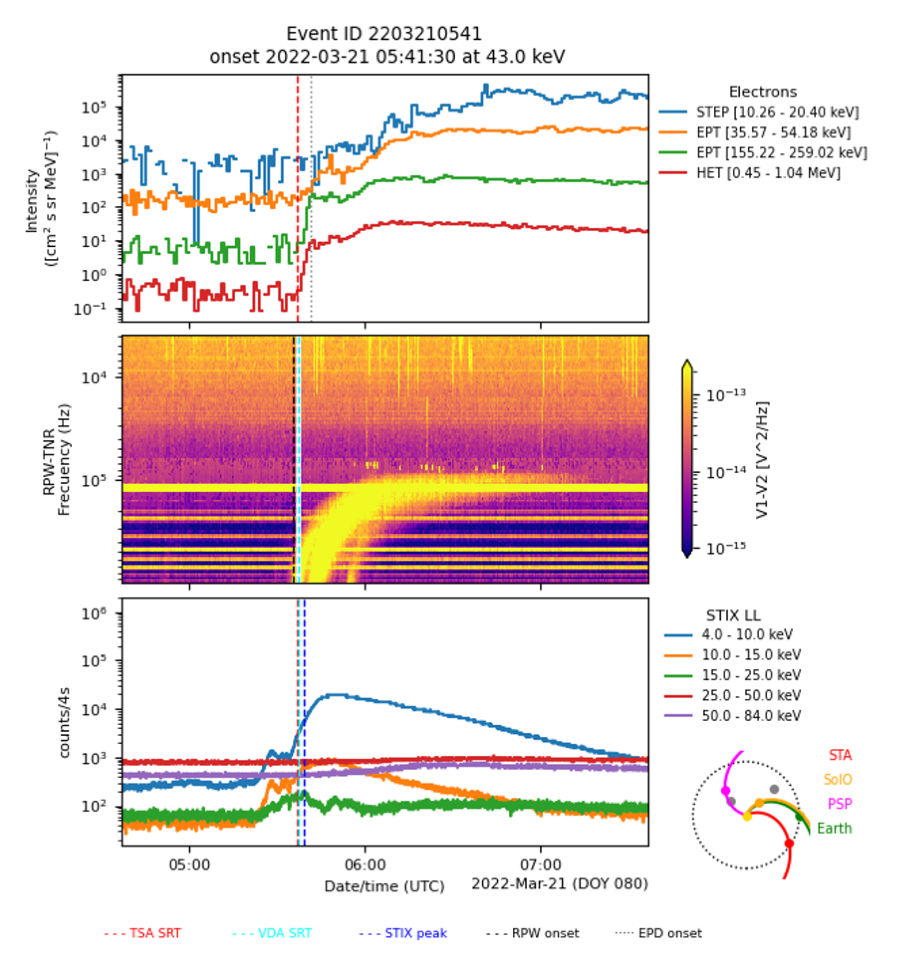}
    \caption{Energetic electrons observed from 10 keV to 1 MeV (top panel), the radio dynamic spectrum showing a type III radio burst close to the electron release time and the X-ray light curves in five different energy bands from 4 to 84 keV. This is one of the many energetic electron events recorded in the Cosee-Cat catalogue \citep{2025A&A...701A..20W}. }
    \label{fig:fig4}
\end{figure}

Non-thermal electrons are efficiently accelerated in solar flares \citep[e.g.][as reviews]{2011SSRv..159..107H,2011SSRv..159..301K} and can escape into the interplanetary space along open field lines. These electrons can be observed remotely as type III solar radio bursts and directly in situ as solar energetic electrons. A key question is whether the same acceleration mechanism produces both the electrons responsible for hard X-ray emission at the Sun and those that escape into the heliosphere. Hard X-rays are often observed co-temporally with type III radio bursts and in situ energetic electrons, as is evident in Fig.~\ref{fig:fig4}, which is one of the many energetic electron events observed with SolO and part of the CoSEE-Cat catalogue \citep{2025A&A...701A..20W}. The correlation between the spectral index of electrons measured at the Sun via hard X-rays and near the Earth measured in situ argues in favour of a common origin of these electrons \citep{2007ApJ...663L.109K,2017MNRAS.471...89J}. The observed correlation between the starting frequencies of type III bursts and the spectral index of X-ray producing electrons \citet{2011A&A...529A..66R,2014A&A...567A..85R} further supports this direct connection.

However, there are a number of inconsistencies and puzzles that show the situation is more complex than a single, instantaneous release of one electron population. Firstly, the number of accelerated electrons at the Sun is typically 2-3 orders of magnitude larger than what is observed at 1~AU \citep{1981AdSpR...1c..73D,2007ApJ...663L.109K, Dresing2021}, implying that only a fraction of the accelerated electrons escape efficiently onto the field lines that connect to the observer or that significant losses occur during transport. Secondly, the timing of type III bursts, hard X-ray emission, and the onset of near-relativistic electrons in situ is not always well-correlated \citep{2002ApJ...579..841H,2004ApJ...609.1082M}. Near-relativistic electrons often arrive minutes later than expected from assuming that the first type III signature marks their solar injection time. This behavior was identified at 1~AU in impulsive electron events \citep{2002ApJ...579..841H} and in studies that compare in situ electron injection with coronal shocks \citep{kl02}. Cases were also found where near-relativistic onsets lagged behind the start time of type III bursts, while radio bursts and X-rays were co-temporal at the Sun \citep{2003ApJ...598.1403C}. Statistical summaries reported delays of several minutes with significant event to event variability \citep{2006GeoRL..33.3106W}. Measurements inside 1~AU confirm that such delays remain common and that the delay varies with heliocentric distance and magnetic connection \citep{2025ApJ...980...96M,2025A&A...701A..20W}. A practical conclusion is that type III timing captures the first escape of suprathermal electrons along particular field lines, while the in situ near-relativistic onsets record a convolution of that escape with transport through a structured medium \citep{2025ApJ...980...96M, 2025A&A...701A..20W}. Another possibility is that the injection is delayed either due to particle trapping or due to the shock that only becomes efficient or magnetically connected later during the eruption \citep[e.g.][]{morosan2024}. Thirdly, not all flares are associated with type III bursts and vice versa \citep{2011A&A...529A..66R}. Finally, the spectrum of electrons observed in situ often shows a broken power-law and even more complex shapes have now been observed thanks to the high energy resolution of SolO's energetic particle detector (see for example Fig.~7 of \citealt{morosan2025}), whereas the electron spectrum responsible for hard X-rays is often a single power-law \citep{2007ApJ...663L.109K,2017MNRAS.471...89J}. Thus, it is essential to determine the location of type III bursts with high accuracy and their relation to hard X-ray sources in order to understand how electrons are released and transported.

Difficulties in connecting type III bursts to solar energetic electrons include the need for precise spatial information together with the time-frequency resolution required to accurately determine the trajectories of the electron beams near their starting frequencies. By comparing large-scale simulations \citep{2001SoPh..202..131K,2013SoPh..285..217R,2014A&A...572A.111R} with the observations, we can identify the processes governing the chief type III solar radio burst characteristics. If the flare electrons detected in situ are produced in, and/or transported through, hot, over-dense regions close to X-ray emitting electrons, then this plasma signature should be present in their lower-energy spectra (1--20~keV), observable in situ \citep{2023ApJ...958...18P}, and possibly in the drift rate of type III bursts as well.

Two aspects explain much of the observed behavior in timing and spectra. The propagation is not stationary at a given height, since the fastest portion of the distribution arrives first and the local population changes rapidly as slower electrons pass \citep{2013SoPh..285..217R,2025ApJ...990..100K}. The background plasma is also randomly inhomogeneous, which modifies the resonance with Langmuir waves and reshapes both the wave spectrum and the electron distribution as they propagate \citep{2001A&A...375..629K,2012ApJ...761..176R,2013ApJ...778..111K, 2013AnGeo..31.1379V,2014JGRA..119.4239B, 2015ApJ...807...38V, 2016ApJ...833..166V, 2015JGRA..12010139V, 2025ApJ...990..100K}. These effects lengthen relaxation scales and weaken a one to one mapping between the first type III signatures and the later arrival of higher energy electrons \citep{2013SoPh..285..217R}. They can also introduce spectral breaks in peak flux versus energy and change the apparent release chronology at different observers. Direct evidence for wave--beam coupling that modifies beam spectra has been reported with inner heliosphere measurements \citep{2023ApJ...959..128L}, and threshold velocities for wave growth have been derived consistently with that picture \citep{2023SoPh..298...52L}. Models that treat the front of the beam as a time-dependent driver of instability in an inhomogeneous plasma further clarify why the earliest radio signatures do not always map directly to the earliest near-relativistic onsets.

In summary, hard X-rays, type III radio bursts and in situ energetic electrons are often clearly related and in many cases consistent with a common origin, but discrepancies in number, timing, association rates and spectral shape show that electron release and subsequent transport through a structured and inhomogeneous corona and heliosphere play a decisive role in what is ultimately observed.

\subsection{Electron density mapping}
\label{section3.4}

The historical remote sensing observations of eclipse and white-light coronagraph observations provided several electron density models \citep{new61, saito77}. Drift rates of type III bursts are also used to derive the coronal electron densities \citep{Leblanc95}. But all these models are mostly one-dimensional (1D) density models. 1D models are mostly based on radial symmetry and they do not account for the dynamic changes caused by the solar transients. Also, they are unable to capture the small-scale density fluctuations and are often scaled to match the in situ measurements. The recent \cite{Kruparova23} model employed high-resolution in situ observations from PSP. However, it should be noted that PSP measures only local densities while traveling at high speed along its orbit. 

Type III bursts are an excellent tool for estimating the coronal electron density. The direct relation between plasma frequency and electron density combined with imaging observations allow mapping of the electron densities along the propagation path of the burst. It is often observed that radio derived densities are usually larger than the densities obtained from density models and multiples of density models are used to match the observations \citep{Pohjolainen08, 2018ApJ...868...79C,jebaraj2020, Reid2020, Deshpande2025}. However, correcting for radiowave scattering allows to obtain the density profiles that are consistent with other measurements without multipliers \citep{2019ApJ...884..122K,2023ApJ...956..112K,2025NatSR..1511335C}. Although many studies also suggest that type III bursts propagate along high-density regions \citep{Trottet1982, Kundu1984, mccauley2018, Reid2020}.  

Accurate mapping of electron density is important to understand the generation of type III bursts, their propagation through ambient plasma and electron beam dynamics. It also provides insights into the density fluctuation level and how it influences the electron beam propagation. SKA in combination with other available ground and space-based observations and using corrections for scattering \citep{2023ApJ...956..112K} will enable detailed mapping of the radio derived density to understand associated physical processes.

\subsection{Continuum radio emission and energetic particles}
\label{section3.5}

Broadband continua provide complementary constraints on the electrons that shape in situ events. Stationary type~IV continua at microwave wavelengths trace trapped non-thermal electrons in post flare or arcade like fields. Microwave imaging spectroscopy retrieves the non-thermal electron distribution and the magnetic field strength and its evolution in those sources. That constrains the number of accelerated electrons that remain confined and the conditions for escape along the open field \citep{Gary2018}. Events with strong and long-lived microwave continua suggest sustained acceleration or efficient trapping, which can delay escape and change the hardness of the seed population that later appears in situ \citep{Fleishman2020}. 

Moving type~IV like continua at metric to decametric wavelengths trace energetic electrons carried outward from the Sun. Imaging and spectroscopy show that these sources can mark confined populations behind the CME or inside CME cores and erupting filaments \citep[e.g.][]{mo19b, vasanth2019, klein2024} and in rare cases these electrons can be trapped inside the CME outer structures \citep{bastian2001, chen2025}. When the outward moving continuum lies on field lines that connect to an observer, the timing of in situ electrons and the evolution of their anisotropy can be read together with the continuum drift to identify when the observer samples trapped populations, freshly released beams, or electrons supplied by the shock \citep{Jebaraj2024}. Forward modeling that couples a realistic coronal MHD background with particle transport and gyrosynchrotron radiation shows that electrons injected into a CME flux rope remain largely confined and produce type~IV like gyrosynchrotron spectra, with the strongest emission at CME flanks and a gradual drift of the spectral peak to lower frequencies as the magnetic field weakens during expansion \citep{Husidic2025}. These results connect continuum morphology to the evolving non-thermal electron distribution that can later feed the in situ population when escape paths open \citep{2022A&A...663A.173K}. However, high-resolution observations of these continua are still rare and so far it has been difficult to make in situ connections.

\subsection{Time domain diagnostics between remote and in situ observations}
\label{section3.6} 

For the interpretation of the origin of in situ particles, the comparison between radio onset and particle release times is central. Velocity dispersion and time shift methods \cite[e.g.][]{Vainio2013}, methods used to infer particle injection times, can bias release times and path lengths of particles if background levels, finite injection duration, or scattering are not treated with care \citep{Lintunen2004}. Uncertainty estimates that reflect these effects improve the reliability of inferred release times \citep{2005ApJ...626.1131S}. Ambiguities are reduced when radio onsets and imaging are combined with multi-point observations and well-connected events to determine which magnetic field lines are used by the escaping beams \citep{Krupar14a, morosan2025}.

Polarization adds a further diagnostic of connectivity and emission mode. \cite{2025ApJ...987L..34P} found that during a post-eruption type~III storm observed near perihelion of PSP, the handedness of the circular polarization flipped at the crossing of the heliospheric current sheet and the fractional polarization reached values near 0.4 at several MHz. This behavior is consistent with fundamental \(\mathcal{O}\)-mode emission \citep{2023ApJ...955L..20J} from beams moving along a well ordered fields post-CME eruption and it indicates minimal depolarizing scattering along the radial path. The same event showed a near constant beam speed near $0.1c$, which supports a stable source of repeated injections over many hours and links the radio timing directly to in situ field polarity and path geometry \citep{2025ApJ...987L..34P}. 

The observation of widespread events, where energetic particles are observed by space missions all around the Sun, still challenge our understanding of potentially extreme particle transport processes \cite[][]{Dresing2012} or alternatively the potential existence of very wide, up to circumsolar shocks \cite[e.g.][]{Dresing2025}. New multi-spacecraft observations have even revealed further potential scenarios for widespread SEP events, such as several distinct particle injections spanning wide helio-longitudinal ranges. Repeating groups of type~III and type~II activity were detected in the event of 17 April 2021, lasting over several tens of minutes and providing electron injections into different directions spanning over more than 100° in helio-longitudes \citep{Dresing2023}. A particular novelty in the study of \cite{Dresing2023} was the usage of circular polarization measurements to determine the connectivity of type~III emitting electron beams, an interpretation which has since found consolidating evidence \citep{2025ApJ...987L..34P}. 

Large-scale topology of the interplanetary magnetic field can also change the mapping between radio timing and in situ onset. When a magnetic cloud is present, first arriving electrons can come sunward along loop legs and the effective path inside the structure can exceed simple geometric estimates. Radio signatures still identify the accelerating shock and the field lines that connect to the observer, but the particle timing must be read with that topology in mind \citep{2016A&A...586A..55D,2025A&A...694A..64R}. Close to the Sun, radio continua from synchrotron emitting electrons have been observed together with particle measurements at traveling shocks. These observations point to efficient electron acceleration at strong quasi parallel shocks and they anchor radio interpretations to measured accelerator properties \citep{2024ApJ...976L...7J}.

\subsection{Gamma rays and shock diagnostics}
\label{section3.8}

Long-duration solar gamma ray emission above 100~MeV often scales with the duration and low frequency extent (< 10~MHz) of associated decametric to hectometric type~II emission \citep{2022CEAB...46....2G, 2021ApJ...915...82G, gopalswamy2025solarcyclevariationsustained}. This supports a picture where CME-driven shocks sustain the production of protons above 300~MeV that later return to the lower atmosphere and produce neutral pion decay photons \citep{Gopalswamy2019}. However, whether particles can return into the low atmosphere is an open question. Downward moving particles should experience magnetic mirroring, so only particles in a very narrow pitch‑angle cone (“loss cone”) can reach the low atmosphere \citep{2026SoPh..301....6V}. Simulations by \citep{2022A&A...658A..23H} estimate that only $<2$\% of shock‑accelerated protons  can precipitate back to the Sun even under favourable conditions. However,  \citet{2019ApJ...879...90D} found that precipitation fractions greater than 10\%  are required in the majority of the events. Other events are consistent with flare-accelerated ions trapped in large loops, which motivates joint use of radio imaging and interplanetary type~II drift to identify particle transport paths \citep{PesceRollins2024}. Surveys indicate that events with sustained gamma rays are frequently associated with fast and wide CMEs, which is consistent with long-lived and efficient shocks \citep{Gopalswamy2018, 2025SoPh..300..133M}. 



\subsection{Ground-based constraints for in situ interpretation}
\label{section3.9}

Inn summary, high-resolution radio imaging and spectroscopy is essential for a number of science cases related to SEPs. Low frequency imaging and spectroscopy resolve multiple coronal sources that produce type~III emission in different magnetic funnels and they track how sources move and split as the coronal field opens \citep{morosan2022b}. Imaging of type~II sources shows where shocks intersect coronal structures that guide escaping beams and where electrons are accelerated most efficiently \citep{morosan2024}. The same approach at smaller heliocentric distances has shown low energy electrons accelerated by the shock intersecting field lines that connected to SolO, with in situ electrons correlated in time and space with the imaged radio sources \citep{morosan2025}. At higher frequencies, microwave imaging spectroscopy retrieves the nonthermal electron distribution and the magnetic field in the low corona. That constrains seed populations and escape conditions for later in situ events \citep{Gary2018}. Those measurements also quantify how the energy content and evolution of the coronal field relate to sustained particle acceleration during flares \citep{Fleishman2020}. Used together, these radio constraints define the geometry and timing that particle models must match when they reproduce longitudinal spread and pitch angle evolution.

\section{The Role of the Square Kilometre Array in SEP studies}

The SKA, with its unrivaled sensitivity, dynamic range, and spectro-polarimetric imaging capabilities for solar radio imaging observations, is expected to improve the study of energetic processes on the Sun, particularly those linked to the generation of SEPs as radio observations are direct signatures of electron accelerations in the solar atmosphere. Continuous, high-resolution, high-fidelity solar radio observations across a wide frequency range using the SKA will potentially bridge the gap between coronal radio diagnostics and in situ particle measurements, enabling comprehensive studies of particle acceleration and transport throughout the heliosphere.

\subsection{Synergy with Space Missions and In Situ Measurements}

The SKA will operate during an era of unparalleled heliophysics coordination, alongside missions such as SolO, PSP, Aditya-L1, and the proposed Helioswarm and Lagrange missions. In situ energetic particle measurements from these missions can be temporally and spatially correlated with SKA radio observations to determine SEP injection times, magnetic connectivity, and propagation characteristics.

Such coordinated observations will allow, for the first time, simultaneous high-temporal, spectral and spatial resolution imaging of the acceleration process (via SKA radio data) and direct sampling of its energetic particle products (via in situ detectors). This synergy will provide stringent tests for SEP transport models, including the effects of turbulence, magnetic field line meandering, and cross-field diffusion.

\begin{table}[h!]
\centering
\small
\caption{Observations for diagnosing energetic particles in the heliosphere}
\begin{tabularx}{\textwidth}{>{\raggedright\arraybackslash}p{2cm} 
                          >{\raggedright\arraybackslash}p{2cm} 
                          >{\raggedright\arraybackslash}p{3.3cm} 
                          >{\raggedright\arraybackslash}p{2.5cm} 
                          >{\raggedright\arraybackslash}p{3.4cm}}
\toprule
\textbf{Observation Type} & \textbf{Frequency / Energy Range} & \textbf{Diagnostic / Physical Parameter} & \textbf{Instruments / Missions} & \textbf{Details} \\
\midrule
Solar radio imaging & 50 MHz – 15 GHz & Electron acceleration sites, shock formation, source locations & SKA, LOFAR, MWA, NRH, GRAPH & High time and spatial resolution; imaging spectroscopy crucial \\
\addlinespace
Radio dynamic spectra & kHz – GHz & Type II, III, IV bursts; CME-driven shocks; electron beams & WIND/WAVES, STEREO/WAVES, PSP/FIELDS & Time-frequency evolution traces electron populations and propagation \\
\addlinespace
In situ energetic particles & keV – GeV & Electron/proton flux, spectra, composition & PSP, Solar Orbiter, ACE, WIND, BepiColombo, STEREO, SOHO, Adityia-1 & Links radio emission to particle flux; temporal correlation; magnetic connectivity association \\
\addlinespace
White-light coronagraphy & 1–30 R$_\odot$ & CME structure, shocks & SOHO/LASCO, STEREO/SECCHI, Solar Orbiter/Metis & Helps locate regions of particle acceleration \\
\addlinespace
EUV / X-ray imaging & 0.1–30 nm & Flare energetics, thermal plasma evolution & SDO/AIA, Hinode/XRT, Solar Orbiter/EUI & Correlation with radio bursts identifies acceleration regions \\
\addlinespace
Heliospheric radio triangulation & kHz – MHz & 3D location of radio sources & STEREO, WIND, PSP, Solar Orbiter & Goniopolarimetry allows triangulating electron beams and CME shocks \\
\addlinespace
Magnetic field measurements & nT & Coronal \& heliospheric B-field & PSP/FIELDS, Solar Orbiter/MAG & Critical for modeling particle propagation and wave-particle interactions \\
\bottomrule
\end{tabularx}
\end{table}

\subsection{Quantitative Diagnostics of Acceleration and Magnetic Fields}

Radio spectroscopic imaging offers quantitative diagnostics of key plasma parameters which are considered relevant to SEP acceleration and transportation \citep[e.g.][]{morosan2025}. From plasma emission observations, the local electron density and its variation can be derived \citep[e.g.][]{new61}, while gyrosynchrotron spectral observations can yield information on magnetic field strength and electron energy spectra. Combining these observations, SKA is expected to help determine the spatial and temporal evolution of various plasma parameters such as density, Aflv\'en speed, Aflv\'eniv Mach numbers, shock geometry, that control particle acceleration efficiency \citep[e.g.][]{kumari2017b, kouloumvakos2021, morosan2025}. High-resolution polarimetric images with unprecedented polarization calibration with SKA will provide valuable information about magnetic field orientation and topology \citep{kumari2017c, Reames1999, 2017MNRAS.471...89J, kumari2019direct, normo2025}. Measurement of circular \citep{2014PhDT.......633S, Kumari2019PhD, 2021AGUFMSH15G2090K, 2022SoPh..297...47M, 2023ApJ...958..181K} and linear polarization \citep{Dey2025} across a wide frequency range will enable magnetic field mapping from the chromosphere to several solar radii \citep{2025JApA...46...51N, 2025A&A...700L...3R}, complementing magnetographic data and advancing our understanding of the magnetic connectivity essential for SEP access to the heliosphere.

\subsection{Future observations with SKA}

SKA observations will facilitate similar scientific studies to those outlined in Section~\ref{section3}, with the advantage of a larger frequency bandwidth for solar radio bursts tracking and increased temporal and spectral resolution. For example, SKA-Low combined with SKA-Mid, can observe Type II and other radio bursts at higher frequencies compared to LOFAR. High-frequency type II bursts, starting at frequencies over 500~MHz have previously been reported \citep{kumari2017a, vasanth2025}, indicating the presence of early low coronal shock waves. The larger observing bandwidth of SKA allows for tracking of the electron beams closer to the erupting region and also over a longer distances. This will determine the possible intersecting regions with open field lines in greater detail. Solar observations with the SKA \citep{2019AdSpR..63.1404N} will also provide high fidelity imaging and spectroscopy of type III bursts near/at Hard X-ray producing sites. The high spatial, spectral and temporal resolution of the SKA will allow us to study the detailed evolution of type III bursts and their connection to HXR producing electrons, as well as the detailed evolution of spike bursts and their connection to CME eruptions.

Recent observations have shown that the in Type II \citep{2025A&A...700A.274K}, Type III \citep{2018ApJ...855L...8M} and noise-storm activity \citep{2024ApJ...975..122M}, the acceleration region can be $\sim$ tens of arc-sec to an arc-min at meter wavelengths. The dense \(u-v\) coverage and distribution of longer baselines will allow us to study these regions with unprecedented spatial resolution and dynamic range. 

{Joint observations between SKA-Low (Australia) and SKA-Mid (South Africa) will be subject to practical constraints arising from their geographical separation. Simultaneous observations are limited to the intersection of solar visibility at the two different sites, resulting in a restricted daily overlap window that varies seasonally. However, coordinated time-continuous observations starting from SKA Low, through facilities in Asia, Europe and North America is possible. LOFAR and other European facilities can provide simultaneous observations with SKA-Mid, in particular. }

{Solar radio bursts generally do not directly image the particle acceleration region itself; instead, coherent radio emission traces the non-thermal electron populations after they escape from the acceleration site and propagate along coronal magnetic field structures. Where source sizes are discussed, we refer to the observed radio-emitting regions rather than to the intrinsic acceleration volume, unless explicitly stated otherwise. }

{The anticipated SKA performance depends on staged deployment and upgrade scenarios (e.g., AA4, AA*, and potential future enhancements), which influence sensitivity, instantaneous bandwidth, baseline distribution, dynamic range, and imaging fidelity. The solar science capabilities highlighted here are therefore dependent on the configuration of the appropriate deployment stage.}

\subsection{Future Prospects and Challenges}

The enormous data volume and technical complexity of solar observation with SKA will require non-standard observation strategies, new analysis techniques, and data-handling approaches. Automated trigger-based observations, intelligent event detection algorithm, dynamic spectro-polarimetric imaging pipelines, and machine-learning-based classification of radio fine structures as well as sophisticated radio frequency interference mitigation techniques will be essential for utilizing SKA's oversubscribed observing time efficiently and managing the high-cadence data stream. These tools are also critical for real-time identification of SEP-related events and for the integration of SKA data into global space weather monitoring systems. 

Some of these developments have already been initiated or demonstrated using SKA pathfinders and precursors, such as LOFAR, MWA, MeerKAT, which have pioneered solar dynamic imaging spectroscopy, non-standard observation strategies, and radio frequency interference mitigation techniques to preserve the solar radio bursts \citep{Breitling2015, Zhang2022, Dey2025b, paircars22b, paircars23, kansabanik_2025}. 
First detailed MeerKAT imaging Spectroscopy 
of a solar flare \citep{2026arXiv260205282L} highlight MeerKAT's ability to detect dilute but hot plasma invisible to Extreme Ultraviolet instruments.
These experiments provide a practical foundation for the data-handling, calibration, and automated analysis that SKA will require at much larger scales.

The SKA’s dense \(uv\) coverage and large field of view will also enable simultaneous imaging of the quiet Sun, CME-driven shocks, and interplanetary disturbances, providing a holistic view of the Sun–heliosphere system. With its long baselines, the SKA will achieve unprecedented spatial resolution, enabling the direct identification and morphological characterization of coronal particle acceleration sites. When combined with numerical modeling and data from multiple spacecraft, SKA observations will play a central role in establishing the physical link between coronal energy release, particle acceleration, and heliospheric particle fluxes.

\section*{Acknowledgments}

We acknowledge the CESRA community website curated by Dr.~Eduard Kontar for providing valuable resources related to solar radio observations. D.E.M. acknowledges the Research Council of Finland project `SolShocks' (grant number 354409). A.K. acknowledges support from the ANRF Prime Minister's Early Career Research Grant (PM-ECRG) Program. I.J. acknowledges the Research Council of Finland project `X-Scale', grant No. 371569. K.D. acknowledges the PhD grant from the Royal Observatory of Belgium and the C1 project UnderRadioSun at KU Leuven. N.D. is grateful for support by the Research Council of Finland (SHOCKSEE, Grant 346902, and AIPAD, Grant 368509). D.E.M. and N.D. acknowledge the SOLER project from the European Union's Horizon Europe research and innovation programme under Grant Agreement No. 101134999. Views and opinions expressed are, however, those of the author(s) only and do not necessarily reflect those of the European Union or HaDEA. Neither the European Union nor the granting authority can be held responsible for them.

\bibliographystyle{abbrvnat-maxbibnames4}
\bibliography{chapter.bib} 

\end{document}